\documentclass[aps,final,notitlepage,oneside,twocolumn,nobibnotes,nofootinbib,
superscriptaddress,noshowpacs,centertags]{revtex4-1}

\usepackage[english]{babel}
\usepackage{graphicx}
\usepackage{latexsym}
\usepackage{amssymb}
\usepackage{amsmath}
\usepackage{float}

\begin{document}

\title{Dark matter around primordial black hole at the radiation-dominated stage}
\author{Yu. N. Eroshenko}\thanks{e-mail: eroshenko@ms2.inr.ac.ru}
\affiliation{Institute for Nuclear Research of the Russian Academy of Sciences, Moscow, Russia}

\date{\today}

\begin{abstract}
The accumulation of dark matter particles near the primordial black holes starts at the radiation-dominated cosmological stage and produces the central density spikes. The spikes can be the bright gamma-ray sources due to dark matter annihilation. We present the self-consistent derivation of the equation of motion of particle in the metrics of primordial black hole immersed into cosmological background. By numerical solution of this equation we find the central dark matter density profile. The density growth is suppressed in the central part of the profile compared with previous calculations.
\end{abstract}

\maketitle 




\section{Introduction}	

The possibility of primordial black holes (PBHs) formation was proposed by Ya.B.~Zeldovich and I.D.~Novikov and by S.~ Hawking, see review in \cite{Car03}. Recently the PBHs attracted new attention because their merge in binaries can explain some of the LIGO/Virgo events. Another possible observational signature of PBHs is the annihilation of dark matter (DM) in the density peaks around PBHs \cite{p28,p38,p17}. This effect puts the strong common constraints on the PBHs and DM \cite{Ero16,Bouetal18,Adaetal19} and, in particular, makes the PBHs of stellar masses and weakly interacting DM particles (WIMPs) are almost incompatible \cite{Adaetal19,Beretal19}.

To calculate the DM annihilation one needs the density profile $\rho(r)$ of DM around PBHs. The profiles were considered in several works\cite{Ero16,Bouetal18,Adaetal19}, but usually with some model assumptions. In this paper we use for the calculation the realistic metrics, corresponding to the PBH in the expanding universe at the radiation-dominated stage, obtained in \cite{BabDokEro18}, and the initial conditions for the DM particles motion after kinetic decoupling from radiation. The resulting $\rho(r)$ in its central part differs from the profiles obtained in previous works.

\section{Metrics and the equation of motion}

Let us write the general form of the spherically symmetric metrics in the curvature coordinates
$ds^2 = e^{\nu}dt^2 - e^{\lambda}dr^2 -r^2d\Omega^2$. The mass of PBH is $M_{\rm PBH}$, gravitational radius $r_g=2M_{\rm PBH}$, where $c=G=1$. Thermalized mixture of photons and relativistic particles is called radiation. In the intermediate 
region  
\begin{equation}
r_g\ll r\ll t
\label{region1}
\end{equation}
(far from PBH's and cosmological horizons) the solution for PBH immersed into cosmological background has the form \cite{BabDokEro18}
\begin{equation}
e^\nu \simeq1-\frac{r_g}{r}+\frac{r^2}{4t^2}, \quad 
e^\lambda \simeq1+\frac{r_g}{r}+\frac{r^2}{4t^2},
\label{nu1}
\end{equation}
up to the higher orders of small parameters. And the self-consistent expressions for the density and velocity of  radiation are
\begin{eqnarray}
\rho_r& =& \frac{3}{32\pi t^2}\left(1+\frac{2 r_g}{r}\right),
\label{rhoappr2}\\
u &=& -\frac{3^{3/2}r_g^2}{2r^2}+\frac{3}{4}\frac{r_g}{t}+\frac{r}{2t}.
\label{uappr2}
\end{eqnarray}
Note that the PBH's metrics prevails at $r<r_{\rm infl}=(r_gt^2)^{1/3}$, but the flow stops ($u=0$) only at 
$r=r_{\rm turn}\simeq3^{1/2}(r_g^2t)^{1/3}\ll r_{\rm infl}$ due to the pressure of radiation.

Before the kinetic decoupling, DM particles are connected with radiation and have the same velocity (\ref{uappr2}). After the kinetic decoupling the DM moves freely. Equation (\ref{uappr2}) is different from the velocity that was used in \cite{Adaetal19} at small $r$. This leads to the modification of the DM $\rho(r)$ in the internal region.
To find the equation of free motion of DM particle in the region (\ref{region1}) we need the Christoffel symbol 
\begin{equation}
\Gamma^1_{00}=\frac{1}{2}\left(\frac{r_g}{r^2}+\frac{r}{2t^2}\right)\left(1+\frac{r_g}{r}+\frac{r^2}{4t^2}\right)^{-1} \simeq\frac{1}{2}\left(\frac{r_g}{r^2}+\frac{r}{2t^2}\right).
\label{gamma100}
\end{equation}
In the dimensionless variables \cite{Adaetal19} $\tau=t/r_g$, $y=r/r_g$ the geodesic equation is\footnote{See the analysis of (\ref{eqmain}) in Appendix.}
\begin{equation}
\frac{d^2y}{d\tau^2}=-\frac{1}{2y^2}-\frac{y}{4\tau^2}.
\label{eqmain}
\end{equation}
It coincides with Eq.~(1) of \cite{Adaetal19}. We obtained the same equation in the self-consistent general relativity approach. The turning point $dy/d\tau=0$ could also be obtained from Hamilton--Jacoby equation, but this is out of scope of this paper.

\section{Dark matter density profile}

Now we study the density spike around PBH. As a particular example we consider WIMPs with masses $m_\chi=100$~GeV and $M_{\rm PBH}=10M_\odot$. From \cite{Bouetal18} we obtain the time of decoupling $t_d=0.003$~s. It was argued in \cite{Adaetal19} that for these parameters the thermal velocities of DM are not important, and the density of DM around PBH can be calculated by putting it to the cosmological one at turn-around radius
\begin{equation}
\rho(r)\simeq\bar\rho(t_{\rm ta}(r_{\rm ta}))=\rho_{\rm eq}\left(\frac{t_{\rm eq}}{\tau_{\rm ta}r_g}\right)^{3/2},
\label{densreal}
\end{equation}
where $\rho_{\rm eq}$ is the DM density at the moment $t_{\rm eq}$ of matter-radiation equality. 

\begin{figure}
	\begin{center}
\includegraphics[angle=0,width=0.49\textwidth]{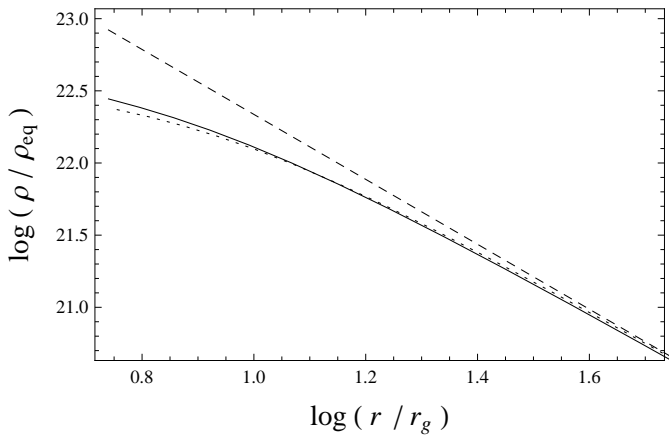}
	\end{center}
\caption{Density of DM around PBH.  Solid curve shows the result of numerical solution of Eq.~(\ref{eqmain}) with initial velocity (\ref{uappr2}). Dashed curve corresponds to the analytic expression (\ref{denssa}), and the dotted curve was obtained from Eq.~(\ref{eqmain}) with the initial velocity $u_i=r_i/(2t_i)$.}
	\label{gr-rho-1}
\end{figure}

The \cite{Adaetal19} found also that the turn-around radius and time are connected by
\begin{equation}
\tau_{\rm ta}\simeq y_{\rm ta}^{3/2},
\label{ta}
\end{equation}
with good accuracy, if the initial conditions are close to the cosmological flow and the second term in (\ref{eqmain}) is initially small in comparison with the first one. Then 
\begin{equation}
\rho(r)=\rho_{\rm eq}\frac{t_{\rm eq}^{3/2}}{r_g^{3/2}}y^{-9/4}
\label{denssa}
\end{equation}
coincides with the famous secondary accretion profile $\rho\propto r^{-9/4}$ \cite{Ber85}, except for the normalisation factor. This density is shown at Fig.~\ref{gr-rho-1} by the dashed curve.

Now we show that the real $\rho(r)$ at small $r$ differs from (\ref{denssa}). We perform the exact numerical solution of  (\ref{eqmain}) with the realistic initial conditions (\ref{rhoappr2}), (\ref{uappr2}) taken at the time of kinetic decoupling $t_d$. We obtain that at small radii the turn-around parameters $\tau_{\rm ta}$ and $y_{\rm ta}$ do not follow the relation (\ref{ta}), and we calculate them numerically. As a result, we obtained from (\ref{densreal}) that the real $\rho(r)$ (solid curve at Fig.~\ref{gr-rho-1}) differs from the calculations of  \cite{Adaetal19}. The density growth in our profile is somewhat suppressed in the center. We had calculated also the $\rho(r)$ with only the last term in the initial condition (\ref{uappr2}), and the result changed slightly. It shows that the exact solution of (\ref{eqmain}) is more important in comparison with the exact initial velocity field.

\section{Conclusion}	

In this paper the equation of motion of free particle was derived self-consistently in the metrics of PBH at cosmological background. The resultant equation coincides with that obtained in \cite{Adaetal19}. By the exact numerical solution of this equation we found the central DM density profile around PBH and demonstrated its central suppression in comparison with profiles from previous works. In the case of WIMPs as DM, this modification does not influence the gamma-ray signal from density spike around PBH because the central part of the DM is totally annihilated and disappeared till now, see the details in \cite{Ero16}. The central $\rho(r)$ survives only for non-annihilating or extremely weakly annihilating DM. Therefore this effect doesn't change the previous constraints on the PBHs and WIMPs.

\appendix

\section{Qualitative study}
\label{appsec}

We can't solve Eq.~(\ref{eqmain}) analytically but we can lower its order and then analyze it qualitatively. It is useful for the understanding the numerical results. Equation (\ref{eqmain}) is homogeneous in general sense, and we use the substitution
$\tau=e^\xi$, $y=w(\xi)e^{m\xi}$, $m=2/3$. Then we lower order by the change $dw/d\xi=p(w)$ and obtain finally
\begin{equation}
p\frac{dp}{dw}+\frac{p}{3}=-\frac{1}{2w^2}-\frac{w}{36}.
\label{aeq1}
\end{equation}
The connection of the old and new variables is the following
\begin{equation}
w=\frac{y}{\tau^{2/3}}, \quad p=\tau^{1/3}\frac{dy}{d\tau}-\frac{2y}{3\tau^{2/3}}.
\end{equation}
It was found in \cite{Adaetal19} by numerical experiments that the turn-around occurs near $y\simeq\tau^{2/3}$ ($w\simeq1$) if the initial condition $\dot y_i=y_i/(2\tau_i)$ was chosen, and  $y_i/(4\tau_i^2)\gg 1/(2y_i^2)$ ($w_i\gg1$). It can be explained by the fast rise of the r.h.s. of (\ref{aeq1}) near $w=1$. Really, consider the $(w,p)$ plane.  If the integral curve of (\ref{aeq1}) begins on the line $\dot y_i=y_i/(2\tau_i)$ at $w_i\gg1$, it has the initial slope $dp/dw=-1/6+6/w_i^3$ and goes almost along the line $p=-w/6$ until $w\simeq1$. Then the curve turns downward. The turn-around points $\dot y=0$ lie at the line $p=-2w/3$ and the integral curve intersect it with $dp/dw=-7/24+3/(4w^3)$ near $w\simeq1$. Therefore, $w\simeq1$ is the attractor of the integral curves with the above initial conditions. This explains the universal relation (\ref{ta}). The different situation will be for $w_i<1$. In this case there is no universal behaviour, and the calculated $\rho(r)$ differs from (\ref{denssa}).

\end{document}